 \documentclass[final,5p,times,twocolumn]{elsarticle}

\usepackage{amssymb}
\usepackage{graphicx}
\usepackage{amsfonts}

\journal{Annals of Physics}

\begin{document}

\begin{frontmatter}



\title{Principles of classical statistical mechanics: \\
A perspective from the notion of complementarity}


\author{Luisberis Velazquez Abad}

\address{Departamento de F\'{i}sica, Universidad Cat\'{o}lica del Norte, Av. Angamos 0610, Antofagasta, Chile.}
\ead{lvelazquez@ucn.cl}
\begin{abstract}
Quantum mechanics and classical statistical mechanics are two physical theories that share several analogies in their mathematical apparatus and physical foundations. In particular, classical statistical mechanics is hallmarked by the \emph{complementarity} between two descriptions that are unified in thermodynamics: (i) the parametrization of the system macrostate in terms of \emph{mechanical macroscopic observables} $I=\left\{I^{i}\right\}$; and (ii) the dynamical description that explains the evolution of a system towards the thermodynamic equilibrium. As expected, such a complementarity is related to the \emph{uncertainty relations} of classical statistical mechanics $\Delta I^{i}\Delta \eta_{i}\geq k$. Here, $k$ is the Boltzmann's constant, $\eta_{i}=\partial \mathcal{S}(I|\theta)/\partial I^{i}$ are the \emph{restituting generalized forces} derived from the entropy $\mathcal{S}(I|\theta)$ of a closed system, which is found in an equilibrium situation driven by certain control parameters $\theta=\left\{\theta^{\alpha}\right\}$. These arguments constitute the central ingredients of a reformulation of classical statistical mechanics from the notion of complementarity. In this new framework, Einstein postulate of classical fluctuation theory $dp(I|\theta)\sim\exp\left[\mathcal{S}(I|\theta)/k\right]dI$ appears as the correspondence principle between classical statistical mechanics and thermodynamics in the limit $k\rightarrow0$, while the existence of uncertainty relations can be associated with the non-commuting character of certain operators.

\end{abstract}


\end{frontmatter}


\section{Introduction}

In physics, complementarity was introduced by Niels Bohr as a basic principle of quantum theory \cite{Bohr0,Heisenberg0}, which refers to effects as the \emph{wave-particle duality}. In an analogous fashion like the finite character of the speed of light $c$ implies the impossibility of a sharp separation between the notions of space and time, the finite character of the quantum of action $\hbar$ implies the impossibility of a sharp separation between the behavior of a quantum system and its interaction with the measuring instruments. The notion of complementarity is closely related to the existence of \emph{complementary quantities}. The best example are the position $\mathbf{q}$ and the momentum $\mathbf{p}$, whose components are related by Heisenberg uncertainty relations:
\begin{equation}\label{HUR}
\Delta q^{i} \Delta p_{i} \geq\hbar/2.
\end{equation}
These inequalities reveal that classical description is limited, in particular, the notion of particle trajectory $\left[\mathbf{q}(t),\mathbf{p}(t)\right]$.

Bohr understood that complementarity cannot be a unique feature of quantum mechanics, but any physical theory with a statistical formulation \cite{Bohr,Heisenberg}. In particular, he suggested that the thermodynamical quantities of temperature $T$ and energy $E$ should be complementary in the same way as position $\mathbf{q}$ and momentum $\mathbf{p}$. According to Bohr, a \emph{definite temperature} $T$ can be only attributed to the system if it is submerged into a \emph{heat bath}, in which case fluctuations of energy $E$ are unavoidable. Conversely, a definite energy $E$ can be only assigned when the system is put into \emph{energetic isolation}, thus excluding the simultaneous determination of its temperature $T$. At first glance, these arguments are remarkably analogous to those ones that justify the complementary character between the position $\mathbf{q}$ and the momentum $\mathbf{p}$. Rosenfeld employed the underlying analogy with quantum mechanics and obtained in the framework of classical fluctuation theory a quantitative uncertainty relation in the form $\Delta E \Delta(1/T)\geq k$ \cite{Rosenfeld}, where $k$ is the Boltzmann's constant. However, this approach was performed under special restrictions which meant that the fluctuations of energy and temperature became dependent on each other and were no longer really complementary. Along the years, other authors proposed different attempts to support the existence of an \emph{energy-temperature complementarity} inspired on Bohr's arguments \cite{Mandelbrot0,Gilmore,Lindhard,Lavenda,Scholg}. The versions of this relation which have appeared in the literature give different interpretations of the uncertainty in temperature $\Delta\left( 1/T\right) $ and often employ widely different theoretical frameworks, ranging from statistical thermodynamics \cite{Landau} to modern theories of \emph{statistical inference} \cite{Fisher,Rao}. Despite of all devoted effort, this work has not led to a consensus in the literature \cite{Uffink}.

Recently \cite{Vel.ETFR,Vel.URSM}, this old question of statistical mechanics was reconsidered in the framework of classical fluctuation theory analyzing the restrictions of the standard procedure to measure thermodynamic quantities of a given system, namely, considering the thermodynamic equilibrium with a measuring instrument (\emph{a thermometer}). This analysis reveals that the uncertainty relation associated with this type of experimental procedure is given by:
\begin{equation}\label{VC.ETU}
\Delta E \Delta\eta\geq k.
\end{equation}
where $\eta=1/T-1/T^{m}$. Here, $T^{m}$ is the internal temperature of the measuring instrument, which is employed for the indirect estimation of the system temperature $T$ via the condition of thermal equilibrium. This new result does not support a complementary character between \emph{thermal contact} and \emph{energetic isolation} as Bohr previously suggested. In fact, this inequality was obtained assuming that a system has \emph{a definite temperature} in conditions of energetic isolation\footnote{Temperature notion referred to in Bohr's argument is the temperature employed in thermal physics, which is the control parameter of the known Gibbs' canonical ensemble $dp(E|T)=\exp(-E/kT)/Z(T)$. Conversely, the temperature notion associated with inequality (\ref{VC.ETU}) is the \emph{microcanonical temperature} $1/T=\partial S(E)/\partial E$, with $S(E)$ being the system entropy. These two temperature notions are equivalent in the framework of \emph{extensive systems} considered in everyday practical applications. However, the difference between these two temperature notions turns crucial in the framework of \emph{nonextensive systems} such as astrophysical systems and small systems, overall, when anomalous states with \emph{negative heat capacities} are present in the thermodynamic description.}. However, the statistical relevance of the temperature implies that its experimental determination is restricted by statistical uncertainties exhibiting a complementary character. Consequently, the system temperature $T$ cannot be regarded as the complementary quantity of the energy $E$, but the \emph{inverse temperature difference} $\eta=1/T-1/T^{m}$. Notice that this quantity can be regarded as the \emph{restituting generalized force} that drives the energetic interchange between the system and the measuring instrument, and hence, the tendency of this closed system towards the thermodynamic equilibrium.

The goal of this work is to deepen on conceptual analogies between quantum theory and classical statistical mechanics. The discussion starts in section \ref{Review} reviewing some important antecedents, such as representations employed in thermodynamics and the general theorems of Einstein fluctuation theory. Afterwards, it will be shown in section \ref{CCSM} that classical statistical mechanics admits a reformulation remarkably analogous to quantum mechanics, which includes the statistical counterparts of \emph{complementarity and correspondence principles}, as well as the reinterpretation of thermodynamic complementary quantities with certain \emph{non-commuting operators}. Finally, section \ref{Discussions} will be devoted to discuss the possible implications of the present reinterpretation, such as the existence of a counterpart of Schr\"{o}dinger equation in the framework of classical fluctuation theory.

\section{Some relevant antecedents}\label{Review}

\subsection{Thermodynamic representations}
Historically, thermodynamics attributes a special preference to the energy $E$ and the temperature $T$ in regard to other thermodynamic variables. This preference is manifested in most of thermodynamic expressions, e.g.: the total differential of the energy \cite{Reichl}:
\begin{equation}\label{Energy}
dE=TdS-Y_{\alpha}dX^{\alpha}.
\end{equation}
Here, $S$ is the entropy, $X^{\alpha}=(V,\mathbf{M},\mathbf{P},N^{s},\ldots)$ are the generalized displacements (volume $V$, magnetization $\mathbf{M}$, polarization $\mathbf{P}$, the numbers of species $N^{s}$, etc.) and $Y_{\alpha}=(p,-\mathbf{H},-\mathbf{E},-\mu_{s})$ the corresponding generalized forces (pressure $p$, magnetic field intensity $\mathbf{H}$, electric field intensity $\mathbf{E}$, the chemical potentials $\mu_{s}$, etc.). Eq.(\ref{Energy}) is said to be written in the \emph{energy representation}. Accordingly, the entropy $S$ and the temperature $T$ are regarded as \emph{conjugated thermodynamic variables}. This type of relation is observed in other thermodynamic potentials, e.g.. the free energy:
\begin{equation}
F=E-TS\rightarrow dF=-SdT-Y_{\alpha}dX^{\alpha},
\end{equation}
which also exhibit an explicit energetic relevance.

Although the energy representation is very useful in practical applications of thermodynamics, the \emph{entropy representation}:
\begin{equation}
dS=\frac{1}{T}\left(dE+Y_{\alpha}dX^{\alpha}\right)\equiv\beta dE+\xi_{\alpha}dX^{\alpha}
\end{equation}
is the most relevant from the viewpoint of \emph{statistical mechanics}. The bedrock this physical theory is the \textit{statistical entropy} introduced by Boltzmann in his attempt to demonstrate the statistical relevance of \textbf{second principle of thermodynamics} \cite{Reichl}:
\begin{equation}\label{Entropy}
S=k\log \Omega.
\end{equation}
Here, $\Omega$ is the number of microstates (microscopic configurations) corresponding to a given macrostate of the thermodynamic system, which is determined by the set of \emph{mechanical macroscopic observables}, e.g.: the energy $E$ and the generalized displacements $X^{\alpha}$, $S=S(E,X)$, which have a direct relevance from microscopic laws of physics. Remarkably, the entropy representation does not attribute any preference to the energy $E$ in regard to other mechanical macroscopic observables $X$. Hence, it is worth introducing the notation conventions:
\begin{equation}
(E,X^{\alpha})\rightarrow I=\left\{I^{i}\right\}\mbox{ and }(\beta, \xi_{\alpha})\rightarrow\mathbf{\zeta}=\left\{\zeta_{i}\right\}.
\end{equation}
This simple consideration allows us to deal with more symmetry and simplicity in mathematical expressions using, as example, the Einstein summation convention of repeated indexes:
\begin{equation}\label{convention}
dS=\beta dE+\xi_{\alpha}dX^{\alpha}\equiv \zeta_{i}dI^{i}.
\end{equation}
While generalized displacements $I$ exhibit \emph{a mechanical relevance} within microscopic laws of physics, the \emph{generalized forces} $\zeta$ exhibit the same \emph{statistical relevance} of the entropy $S$.

\subsection{Einstein fluctuation theory}

Classical fluctuation theory starts from \textbf{Einstein postulate} \cite{Landau}:
\begin{equation}\label{EP}
dp(I|\theta)=C_{\theta}e^{\frac{1}{k}\mathcal{S}(I|\theta)}dI,
\end{equation}
which describes the fluctuating behavior of a set of macroscopic observables $I=\left\{I^{i}\right\}$ in an equilibrium situation driven by certain set of control parameters $\theta=\left\{\theta^{\alpha}\right\}$. Here, $\mathcal{S}(I|\theta)$ is the entropy of a closed system corresponding to a macrostate $(I|\theta)$, and $C_{\theta}$ is a normalization factor that depends on control parameters $\theta$. Einstein introduces hypothesis (\ref{EP}) rephrasing the statistical entropy (\ref{Entropy}) as $\Omega(I|\theta)=e^{\frac{1}{k}\mathcal{S}(I|\theta)}$ to obtain \emph{relative probabilities}:
\begin{equation}
dp(I|\theta)=\frac{e^{\frac{1}{k}\mathcal{S}(I|\theta)}
dI}{\int_{\mathcal{M}_{\theta}} e^{\frac{1}{k}\mathcal{S}(I|\theta)}dI}\equiv C_{\theta}e^{\frac{1}{k}\mathcal{S}(I|\theta)}dI.
\end{equation}
Here, $\mathcal{M}_{\theta}$ denotes the abstract space constituted by all admissible values of set of macroscopic observables $I$ that are accessible for a fixed $\theta\in\mathcal{P}$, where $\mathcal{P}$ is the space of all admissible values of control parameters $\theta$ of the closed system. Conventionally, classical fluctuation theory is applied to describe the small fluctuating behavior observed in large thermodynamic systems, which are analyzed in the framework of the \emph{gaussian approximation} \cite{Landau}. This description is obtained expanding the entropy $\mathcal{S}(I|\theta)$ about its equilibrium value as follows:
\begin{equation}
\mathcal{S}(I|\theta)=\mathcal{S}(\bar{I}|\theta)+
\frac{1}{2}\frac{\partial^{2}\mathcal{S}(\bar{I}|\theta)}{\partial I^{i}\partial I^{j}}\alpha^{i}\alpha^{j}+O\left(\alpha^{2}\right).
\end{equation}
Here, $\alpha^{i}=I^{i}-\bar{I}^{i}$ is the deviation from the \emph{equilibrium configuration} $\bar{I}=\bar{I}(\theta)$:
\begin{equation}\label{equilibrium}
\frac{\partial \mathcal{S}(\bar{I}|\theta)}{\partial I^{i}}=0\mbox{ and } -\frac{\partial^{2}\mathcal{S}(\bar{I}|\theta)}{\partial I^{i}\partial I^{j}}\succ 0,
\end{equation}
where the notation $A_{ij}\succ 0$ indicates that the matrix $A_{ij}$ is positive definite. Introducing the matrix $\bar{g}_{ij}(\theta)$ given by the negative of the entropy Hessian at the equilibrium configuration:
\begin{equation}\label{Hess}
\bar{g}_{ij}(\theta)=-\frac{\partial^{2}\mathcal{S}(\bar{I}|\theta)}{\partial I^{i}\partial I^{j}},
\end{equation}
the original distribution function (\ref{EP}) can be approximated as:
\begin{equation}\label{gaussian}
dp_{G}(\alpha|\theta)=\sqrt{ \frac{\det\left|\bar{g}_{ij}(\theta)\right|}{(2\pi k)^{n}}}e^{-\frac{1}{2k}\bar{g}_{ij}(\theta)\alpha^{i}\alpha^{j}}d\alpha.
\end{equation}
Here, $\det\left|\bar{g}_{ij}(\theta)\right|$ denotes the determinant of the matrix $\bar{g}_{ij}(\theta)$, with $n$ being the dimension of the manifold $\mathcal{M}_{\theta}$, that is, the number of mechanical macroscopic observables $I$. The self-correlation matrix $\left\langle\alpha^{i}\alpha^{j}\right\rangle$ can be estimated in the gaussian  approximation (\ref{gaussian}) by the inverse $\bar{g}^{ij}(\theta)$ of the matrix (\ref{Hess}):
\begin{equation}\label{Hess.fluct}
\left\langle\alpha^{i}\alpha^{j}\right\rangle=k\bar{g}^{ij}(\theta).
\end{equation}

\subsection{Restituting generalized forces}

The entropy of the closed system $\mathcal{S}(I|\theta)$ allows to introduce the \emph{restituting generalized forces} $\eta_{i}$:
\begin{equation}\label{dgf}
\eta_{i}(I|\theta)=\frac{\partial \mathcal{S}(I|\theta)}{\partial I^{i}}.
\end{equation}
Geometrically, these quantities represent the components of a vector field $\eta=\left\{\eta_{i}\right\}$ defined on the space $\mathcal{M}_{\theta}$, which is oriented towards the equilibrium configurations $\bar{I}$. For a particular illustration, let us consider the closed system composed of a system of interest and its environment (e). Most of practical applications of thermodynamics are focussed on huge systems driving by short-range forces, commonly referred to as \emph{extensive systems}, whose relevant macroscopic observables obey an additive constraint $\theta=I+I^{e}$ (e.g.: energy, volume, the number of chemical species, etc.) and the entropy $\mathcal{S}(I|\theta)$ is decomposed as follows $\mathcal{S}(I|\theta)=\mathcal{S}(I)+\mathcal{S}^{e}(I^{e})$. Thus, the restituting generalized forces $\mathbf{\eta}$ associated with this type situations are simply the difference of the generalized forces $\zeta$ and $\zeta^{e}$ of the system and its environment, $\eta=\zeta-\zeta^{e}$. Moreover, the vanishing of each component $\eta_{i}$ at the equilibrium configurations $\bar{I}$ drops to well-known \emph{thermodynamic equilibrium conditions},  $\bar{\zeta}_{i}=\bar{\zeta}^{e}_{i}$. In non-equilibrium thermodynamics, the restituting generalized forces $\mathbf{\eta}$ appear in the \textit{hydrodynamic equations} describing the system relaxation towards the equilibrium. Usually, such equations exhibit the following phenomenological form:
\begin{equation}\label{transport}
\frac{d}{dt}I^{i}(t)=L^{ij}\eta_{j}(t)
\end{equation}
in the framework of linear transport theory \cite{Reichl}. Here, $L^{ij}$ is the matrix of \emph{transport coefficients}, which obeys the known \emph{Onsager's conditions} $L^{ij}=L^{ji}$.

\subsection{Rigorous theorems in Einstein fluctuation theory}

As naturally expected, gaussian approximation of Einstein fluctuation theory has a restricted applicability. Even in the framework of extensive systems, gaussian distribution (\ref{gaussian}) is inappropriate to describe their fluctuating behavior in certain situations. For example, the exact distribution function (\ref{EP}) can exhibit a \emph{multimodal character} for certain values of control parameters $\theta$, that is, the existence of two or more (meta)stable equilibrium configurations $\bar{I}$. Moreover, some components of the matrix $\bar{g}^{ij}(\theta)$ can undergo a divergence for certain values of control parameters $\theta$, so that, gaussian distribution (\ref{gaussian}) can overestimate in a significant way the system fluctuating behavior. The first situation is typical during the occurrence of discontinuous phase transitions, while the second one takes place at the critical points of continuous phase transitions \cite{Reichl}.

Fortunately, the previous inconveniences can be overcome. Under general mathematical conditions, Einstein postulate (\ref{EP}) allows to obtain some rigorous fluctuation theorems expressed in term of the restituting generalized forces (\ref{dgf}):
\begin{equation}\label{correlation}
\left\langle\eta_{i}\right\rangle=0, \:
\left\langle \delta I^{j}\eta_{i}\right\rangle=-k\delta^{j}_{i}, \:
\left\langle -\partial\eta_{j}/\partial I^{i}\right\rangle=\left\langle \eta_{i}\eta_{j}\right\rangle.
\end{equation}
These theorems were recently derived by Velazquez and Curilef in an attempt to arrive at a set of equilibrium fluctuation relations compatible with anomalous response functions, e.g.: \emph{negative heat capacities} \cite{Vel.GEFT}. The same ones have a paramount importance in the present work, so that, let us briefly commented their physical relevance.

The first fluctuation theorem is a statistical form of \textbf{conditions of thermodynamic equilibrium}. Such a physical relevance is easy to verify taking into consideration the closed system referred to as a particular example in the previous subsection. As already commented, the restituting generalized forces $\eta_{i}$ for this situation are given by the difference among the generalized forces of the system and its environment, $\eta_{i}\equiv\zeta_{i}-\zeta^{e}_{i}$, and therefore, $\left\langle\eta_{i}\right\rangle\equiv 0\rightarrow\left\langle\zeta_{i}\right\rangle=\left\langle\zeta^{e}_{i}\right\rangle$.

The second fluctuation theorem states that the fluctuations of the $i$-th mechanical observable $\delta I^{i}$ are only correlated to its conjugated restituting generalized force $\eta_{i}$. Physically, the second fluctuation theorem is a statistical form of the \textbf{Le Ch\^{a}telier principle} \cite{Reichl}: \emph{the existence of a spontaneous fluctuation $\delta I^{i}$ will provoke the existence of a generalized force $\eta_{i}$ that restitutes the system towards its equilibrium configuration}. Interestingly, the correlation function $\left\langle \delta I^{i}\eta_{i}\right\rangle$ exhibits a relevant constant value: the negative of the Boltzmann's constant $k$, which is the natural unit of the statistical entropy (\ref{Entropy}). At first glance, this result reinforces the \emph{conjugated character} between the mechanical observable $I^{i}$ and the corresponding restituting generalized force $\eta_{i}$. Noteworthy that this type of relationship cannot be obtained for the ``\emph{conjugated variables}'' of the energy representation ($S\leftrightarrow T$ and $X^{\alpha}\leftrightarrow Y_{\alpha}$). In fact, correlation functions as $\left\langle\Delta S\Delta T\right\rangle$ and $\left\langle\Delta X^{\alpha}\Delta Y_{\alpha}\right\rangle$ are expressed into \emph{energy units}, and hence, the same ones must correspond to certain thermodynamics functions instead of a universal physical constant like the Boltzmann's one $k$.

Finally, the third fluctuation theorem is a statistical form of the \textbf{conditions of thermodynamic stability}. Since the self-correlation function of the restituting generalized forces $\left\langle\eta_{i}\eta_{j}\right\rangle$ is always a positive definite matrix, the statistical expectation value of the \emph{response matrix} $\chi_{ij}=-\partial\eta_{j}/\partial I^{i}$ of the closed system is also a positive definite matrix, $\left\langle\chi_{ij}\right\rangle\succ 0$. Physically speaking, the thermodynamic stability conditions contain the same qualitative information of the Le Ch\^{a}telier principle.

\section{Principles of classical statistical mechanics}\label{CCSM}

\subsection{Uncertainty principle}

A basic assumption of the thermodynamical description is that the conjugated quantities as the mechanical macroscopic observables $I$ and the restituting generalized forces $\mathbf{\eta}$ simultaneously exhibit \emph{definite values} for a given closed system. In particular, this feature is explicitly observed in the hydrodynamic equations (\ref{transport}), where it is possible to speak about a \emph{relaxation curve} $[I(t),\mathbf{\eta}(t)]$ along the system evolution. From the viewpoint of classical statistical mechanics, this type of description is limited. Using the known \emph{Cauchy-Schwartz inequality}:
\begin{equation}\label{Cauchy-Schwartz}
\left\langle A^{2}\right\rangle\left\langle B^{2}\right\rangle\geq \left\langle A B\right\rangle^{2},
\end{equation}
and introducing the \emph{statistical uncertainty} $\Delta x =\sqrt{\left\langle \delta x^{2}\right\rangle}$, the statistical form of the Le Ch\^{a}telier principle (the second fluctuation theorem in Eq.(\ref{correlation})) leads to the following inequality:
\begin{equation}\label{canonical.uncertainty}
\Delta I^{i}\Delta \eta _{i}\geq k,
\end{equation}
which was previously derived in Ref.\cite{Vel.URSM}. Accordingly, the thermodynamic quantities $I^{i}$ and $\eta_{i}$ are not only conjugated, but they are \emph{complementary}. As expected, the energy-temperature uncertainty (\ref{VC.ETU}) is simply a particular case of these inequalities. Rephrasing Max Born's words \cite{Born}, the \textbf{uncertainty principle of classical statistical mechanics} can be enunciated as follows: \emph{There exist a fundamental limit on the accuracy with certain pairs of thermodynamic complementary quantities of a closed system, such as mechanical macroscopic observables $I$ and the restituting generalized forces $\mathbf{\eta}$, can be simultaneously known. In other words, the more precisely one of these quantities is measured, the less precisely the other can be controlled, determined, or known. In consequence, the thermodynamic description is limited}.

Due to the mechanical relevance of each macroscopic observable $I^{i}$, one can obtain a \emph{definite value} of this quantity as result of a single measuring event using an appropriate instrument. Conversely, the statistical relevance of each component of the restituting generalized forces $\eta_{i}$ implies that their determination must involve the \emph{statistical processing} of a large number of the previous measuring events, which should be employed to estimate the probability distribution $dp(I|\theta)$. In particular, one should obtain a good estimation of the statistical entropy $\mathcal{S}(I|\theta)$ in a finite region of the space $\mathcal{M}_{\theta}$ sufficiently large to evaluate the quantities $\eta_{i}$ via definition (\ref{dgf}). At first glance, this requirement is analogous to the determination of the energy $E$ and the momentum $\mathbf{p}$ in quantum mechanics, which demand the knowledge of the wave function $\Psi(\mathbf{q},t)$ (or its square modulus $\left|\Psi(\mathbf{q},t)\right|^{2}$) in a finite region of the space-time $\mathcal{M}^{4}$ sufficient for the determination of the wave frequency $\omega$ and the wave vector $\mathbf{k}$ via \emph{de Broglie's relations} $E=\hbar \omega$ and $\mathbf{p}=\hbar \mathbf{k}$.

Quantum mechanics and classical statistical mechanics are two physical theories that exhibit several analogies in their mathematical aspects and physical foundations, which are briefly summarized in Table \ref{ComparisonQM-CSM}. Both classical mechanics and thermodynamics assume a simultaneous definition of conjugated variables like position $\mathbf{q}$ and momentum $\mathbf{p}$ or the mechanical macroscopic observables $I$ and the restituting generalized forces $\mathbf{\eta}$. However, a different situation is found in those applications where the relevant constants as the quantum of action $\hbar$ or the Boltzmann's constant $k$ are not so small. Accordingly, classical mechanics only provides a very precise description for systems with large quantum numbers, that is, in the limit $\hbar\rightarrow 0$. Analogously, thermodynamics appears as a suitable treatment for systems with a large number $N$ of degrees of freedom, that is, in the limit $k\rightarrow 0$. Otherwise, the curve of the hydrodynamic relaxation $[I(t),\mathbf{\eta}(t)]$ of a closed system is badly defined as the notion of particle trajectory $[\mathbf{q}(t), \mathbf{p}(t)]$. As clearly evidenced, Planck's constant $k$ can be regarded in the framework of classical statistical mechanics as the \emph{quantum of entropy}. Noteworthy that this interpretation was considered in the past by other investigators, as example, Bekenstein employed this argument to derive its famous formula for the black hole entropy \cite{Bekenstein}.

\begin{table*}[tbp] \centering%
\begin{tabular}{||c|c|c||}
\hline\hline
\begin{tabular}{c}
\textbf{Criterium}
\end{tabular}
& \multicolumn{1}{||c|}{\textbf{Quantum Mechanics}} & \multicolumn{1}{||c||}{%
\begin{tabular}{c}
\textbf{Clas. Stat. Mechanics}
\end{tabular}
} \\ \hline\hline
parameterization &
\begin{tabular}{c}
space-time \\
coordinates $\left( \mathbf{q},t\right) $%
\end{tabular}
&
\begin{tabular}{c}
mechanical macroscopic \\
observables $I $%
\end{tabular}
\\ \hline
\begin{tabular}{c}
probabilistic \\
description%
\end{tabular}
&
\begin{tabular}{c}
The wave function $\Psi \left( \mathbf{q},t\right) $ \\
$dp\left( \mathbf{q},t\right) =\left\vert \Psi \left( \mathbf{q},t\right)
\right\vert ^{2}d\mathbf{q}$%
\end{tabular}
&
\begin{tabular}{c}
The probability amplitude $\Phi \left( I|\theta \right) $ \\
$dp\left( I|\theta \right) =\Phi ^{2}\left( I|\theta \right) dI$%
\end{tabular}
\\ \hline
deterministic theory &
\begin{tabular}{c}
\textbf{classical mechanics} \\
in the limit $\hbar \rightarrow 0$%
\end{tabular}
&
\begin{tabular}{c}
\textbf{thermodynamics} \\
in the limit $k\rightarrow 0$%
\end{tabular}
\\ \hline
\begin{tabular}{c}
relevant physical \\
hypothesis%
\end{tabular}
&
\begin{tabular}{c}
Correspondence principle:  \\
$\Psi \left( \mathbf{q},t\right) \sim e^{\frac{i}{\hbar }S\left( \mathbf{q}%
,t\right) }$ when $\hbar \rightarrow 0$, \\
where $S\left( \mathbf{q},t\right) $ is the action%
\end{tabular}
&
\begin{tabular}{c}
Einstein postulate:  \\
$\Phi \left( I|\theta \right) \sim e^{\frac{1}{2k}S\left( I|\theta \right) }$
when $k\rightarrow 0$, \\
where $S\left( I|\theta \right) $ is the entropy%
\end{tabular}
\\ \hline
evolution & dynamical conservation laws &
\begin{tabular}{c}
tendency towards \\
thermodynamic equilibrium%
\end{tabular}
\\ \hline
\begin{tabular}{c}
conjugated \\
variables%
\end{tabular}
&
\begin{tabular}{c}
energy $E=\partial S\left( \mathbf{q},t\right) /\partial t$ \\
momentum $\mathbf{p}=\partial S\left( \mathbf{q},t\right) /\partial \mathbf{q%
}$%
\end{tabular}
&
\begin{tabular}{c}
restituting generalized forces \\
$\mathbf{\eta}=\partial S\left( I|\theta \right) /\partial I$%
\end{tabular}
\\ \hline
\begin{tabular}{c}
complementary \\
quantities%
\end{tabular}
& $\left( \mathbf{q},t\right) $ \textit{versus} $\left( \mathbf{p},E\right) $
& $I$ \textit{versus} $\mathbf{\eta}$ \\ \hline
\begin{tabular}{c}
operator \\
representation%
\end{tabular}
& $\hat{q}^{i}=q^{i}$ and $\hat{p}_{i}=-i\hbar \frac{\partial }{\partial
q^{i}}$ & $\hat{I}^{i}=I^{i}$ and $\hat{\eta}_{i}=2k\frac{\partial }{%
\partial I^{i}}$ \\ \hline
\begin{tabular}{c}
commutation \& \\
uncertainty relations%
\end{tabular}
& $\left[ \hat{q}^{i},\hat{p}_{j}\right] =i\hbar \delta _{j}^{i}\Rightarrow
\Delta q^{i}\Delta p_{i}\geq \hbar /2$ & $\left[ \hat{I}^{i},\hat{\eta}_{j}%
\right] =-2k\delta _{j}^{i}\Rightarrow \Delta I^{i}\Delta \eta _{i}\geq k$
\\ \hline\hline
\end{tabular}%
\caption{Comparison between quantum mechanics and classical statistical
mechanics. Despite their different physical relevance, these theories exhibit several analogies as consequence of their statistical nature.}
\label{ComparisonQM-CSM}
\end{table*}

\subsection{Complementarity principle}

Quantum mechanics is hallmarked by the complementarity between two descriptions that are unified in classical mechanics: (i) the space-time description, that is, the parametrization of the system state in terms of the position $\mathbf{q}$ and the time $t$; and (ii) the dynamical description based on the conservation laws. Analogously, classical statistical mechanics is hallmarked by the complementarity between two descriptions that are unified in thermodynamics: (i) the parametrization in terms of mechanical macroscopic observables $I$; and (ii) the dynamical description that explains the tendency of the system towards the thermodynamic equilibrium. From the mathematical viewpoint, uncertainty relations appear as a consequence of the \emph{coexistence of variables with different relevance} in a physical theory with a statistical formulation. In one hand, there exist the variables that parameterize the \emph{existence} of a closed system, such as the space-time coordinates $(t,\mathbf{q})$ or the mechanical macroscopic observables $I$. In the other hand, there exist the conjugated variables like the energy $E$ and the momentum $\mathbf{p}$ or the restituting generalized forces $\mathbf{\eta}$, which only exhibit a statistical relevance in the respective dynamical descriptions.

As expected, the emergence of complementarity in a closed system occurs when one attempts to measure its properties. Any measuring instrument to study quantum mechanics is a system that obeys classical mechanics with a sufficient accuracy, e.g.: a photographic plate. Analogously, any measuring instrument in classical statistical mechanics is a system that exhibits a well-defined thermodynamical description, e.g.: a thermometer should exhibit a well-defined temperature dependence of its thermometric quantity. As expected, every measuring event involves an uncontrollable perturbation of the system state, which can be significant when the system is sufficiently small. No one measuring instrument can be employed to perform a simultaneous estimation of complementary properties like the conjugated variables $\mathbf{q}\leftrightarrow\mathbf{p}$ and $I\leftrightarrow\mathbf{\eta}$. For example, the experimental determination of the position $\mathbf{q}$ and the volume $V$ demand instruments with rigid scales, while the momentum $\mathbf{p}$ and the generalized forces associated with a pressure gradient $\Delta p$ demand measuring instruments with movable parts. According to uncertainty relations (\ref{HUR}) and (\ref{canonical.uncertainty}), the finite character of the universal physical constants $\hbar$ and $k$ characterizes how much the accuracy achieved during the determination of one of these properties affect the accuracy in the determination of their complementary counterparts.  Rephrasing Bohr's original statement about the quantum complementary principle, the \textbf{complementarity principle of classical statistical mechanics} can be enunciated as follows: \emph{The finite character of the Boltzmann's constant $k$ implies the impossibility of a sharp separation between the behavior of a thermodynamical system and its interaction with the measuring instruments}.

\subsection{Real probability amplitude $\Phi(I|\theta)$}

Quantum mechanics is formulated in terms of \emph{complex probability amplitudes}, that is, \emph{the wave function} $\Psi(\mathbf{q},t)$;
whose square modulus $\left|\Psi(\mathbf{q},t)\right|^{2}$  provides the probability density to detect a particle at the position $\mathbf{q}$ as result of a measuring event at the instant $t$:
\begin{equation}
dp(\mathbf{q},t)=\left|\Psi(\mathbf{q},t)\right|^{2}d\mathbf{q}.
\end{equation}
On the other hand, classical statistical mechanics is a \emph{classical probability theory}, which describes the statistical behavior of a set of real physical quantities in terms of distribution functions:
\begin{equation}
dp(I|\theta)=\rho(I|\theta)dI
\end{equation}
with a \emph{non-negative probability density}, $\rho(I|\theta)\geq 0$. Nevertheless, any classical probability theory can also be rephrased in terms of a \emph{real probability amplitude} $\Phi(I|\theta)$:
\begin{equation}\label{representation}
dp(I|\theta)=\Phi^{2}(I|\theta)dI,
\end{equation}
which is defined on a Hilbert space $\mathcal{H}_{R}$ of real functions. The notion of real probability amplitude can be employed to introduce a scalar product $\left(\cdot\right)$ between two distributions $\Phi^{(1)}=\Phi(I|\theta_{1})$ and $\Phi^{(2)}=\Phi(I|\theta_{2})\in\mathcal{H}_{R}$:
\begin{equation}
\left(\Phi^{(1)}\cdot\Phi^{(2)}\right)\equiv\int_{\mathcal{M}_{\theta}} \Phi(I|\theta_{1})
\Phi(I|\theta_{2})dI.
\end{equation}
This scalar product has been employed in statistical theory to introduce the \emph{statistical distance} $d\left(\Phi^{(1)},\Phi^{(2)}\right)$ between two normalized distribution functions:
\begin{equation}
d\left(\Phi^{(1)},\Phi^{(2)}\right)=\cos^{-1}\left(\Phi^{(1)}\cdot\Phi^{(2)}\right).
\end{equation}
Considering the asymptotic case $\theta_{1}=\theta+\frac{1}{2}d\theta$ and $\theta_{2}=\theta-\frac{1}{2}d\theta$, the previous statistical distance drops to the Riemannian metric of \emph{inference theory} \cite{Rao,Amari}:
\begin{equation}\label{inf.geo}
ds^{2}=\mathfrak{g}_{\alpha\beta}(\theta)d\theta^{\alpha}d\theta^{\beta},
\end{equation}
where $\mathfrak{g}_{\alpha\beta}(\theta)$ is the \emph{Fisher's inference matrix} \cite{Fisher}:
\begin{equation}
\mathfrak{g}_{\alpha\beta}(\theta)=\int_{\mathcal{M}_{\theta}}
\frac{\partial\log\rho(I|\theta)}{\partial\theta^{\alpha}}
\frac{\partial\log\rho(I|\theta)}{\partial\theta^{\beta}}
\rho(I|\theta)dI.
\end{equation}
Wootters employed the analogy with quantum theory to extend the notion of statistical distance in this framework \cite{Wootters}:
\begin{equation}
d\left(\Psi^{(1)},\Psi^{(2)}\right)=\cos^{-1}\left|\left(\Psi^{(1)}\cdot\Psi^{(2)}\right)\right|,
\end{equation}
where $\left(\Psi^{(1)}\cdot\Psi^{(2)}\right)$ is the \emph{hermitian product}:
\begin{equation}
\left(\Psi^{(1)}\cdot\Psi^{(2)}\right)\equiv\int \Psi^{*}_{1}(\mathbf{q},t)\Psi_{2}(\mathbf{q},t)d\mathbf{q}.
\end{equation}

\subsection{Correspondence principle}

Comparing the relations between conjugated mechanical variables $(\mathbf{q},\mathbf{p})$ and the thermodynamical variables $(I,\eta)$:
\begin{equation}
\mathbf{p}=\frac{\partial S(\mathbf{q},t)}{\partial \mathbf{q}}\mbox{ and }\eta=\frac{\partial \mathcal{S}(I|\theta)}{\partial I}
\end{equation}
is evident that the classical action $S(\mathbf{q},t)$ of the Hamilton-Jacobi theory \cite{Esposito} and the statistical entropy $\mathcal{S}(I|\theta)$ appear as two counterpart functions. Correspondence principle of quantum mechanics states the way that quantum description turns equivalent to the classical description in the limit $\hbar\rightarrow 0$. Quantitatively, the same expresses the asymptotic relation between the wave function $\Psi(\mathbf{q},t)$ and the classical action $S(\mathbf{q},t)$:
\begin{equation}\label{CorrespPrinc}
\Psi(\mathbf{q},t)\sim\exp\left[iS(\mathbf{q},t)/\hbar\right]
\end{equation}
in the \emph{quasi-classic limit} $S(\mathbf{q},t)\gg\hbar$. Analogously, the \textbf{correspondence principle of classical statistical mechanics} states the way that classical statistical mechanics turns equivalent to the thermodynamic description in the limit $k\rightarrow0$: \emph{The probability amplitude $\Phi(I|\theta)$ associated with a closed system with a large number of degrees of freedom asymptotically behaves as follows}:
\begin{equation}\label{EP2}
\Phi(I|\theta)\sim \exp\left[\mathcal{S}(I|\theta)/2k\right],
\end{equation}
\emph{where $\mathcal{S}(I|\theta)$ is the thermodynamic entropy.}

As expected, the mathematical form of the asymptotic probability amplitude (\ref{EP2}) has been chosen to match with Einstein postulate (\ref{EP}) of classical fluctuation theory. Accordingly, Einstein postulate (\ref{EP}) should be regarded as an \emph{asymptotic formula} in the thermodynamic limit $\mathcal{S}(I|\theta)\gg k$. Noteworthy that this claim does not contradict conventional applications of classical fluctuation theory, which refer to large thermodynamic systems with a small fluctuating behavior. It has always claimed that quantum mechanics occupies a unusual place among physical theories: it contains classical mechanics as a limiting case, yet at the same time it requires this limit for its own formulation. However, this is not a unique feature of quantum mechanics. Classical statistical mechanics also contains thermodynamics as a limiting case, yet at the same time it requires requires thermodynamic notions for its own formulation.

\subsection{Operators of physical observables}

A relevant hypothesis in the mathematical apparatus of quantum mechanics is that the correspondence of physical observables with certain \emph{Hermitian linear operators}. In particular, the expectation value of a physical observable $O$ is obtained from the rule:
\begin{equation}
\left\langle O\right\rangle=\int \Psi^{*}(\mathbf{q},t)\hat{O}\Psi(\mathbf{q},t)d\mathbf{q},
\end{equation}
where the operator $\hat{O}$ obeys the Hermitian condition:
\begin{equation}
\hat{O}^{+}=\hat{O}.
\end{equation}
Formally, the introduction of quantum operators can be based on the correspondence with classical mechanics. This is the case of the energy and the momentum operators:
\begin{equation}\label{QOperators}
\hat{E}=i\hbar\frac{\partial}{\partial t}\mbox{ and }\hat{\mathbf{p}}=-i\hbar\frac{\partial}{\partial \mathbf{q}}.
\end{equation}
Considering the asymptotic wave function (\ref{CorrespPrinc}), one obtains the following relation:
\begin{equation}
\hat{\mathbf{p}}\Psi(\mathbf{q},t)\sim\hat{\mathbf{p}}e^{i\frac{1}{\hbar}S(\mathbf{q},t)}\rightarrow \mathbf{p}=\frac{\partial S(\mathbf{q},t)}{\partial \mathbf{q}}.
\end{equation}
Thus, the \emph{non-commutating} character between position operators $\hat{\mathbf{q}}=\mathbf{q}$ and momentum operators $\hat{\mathbf{p}}$:
\begin{equation}
\left[\hat{q}^{i},\hat{p}_{j}\right]=i\hbar\delta_{j}^{i}
\end{equation}
can be related to the Heisenberg uncertainty relations (\ref{HUR}) via the Robertson-Schrodinger inequality \cite{Robertson,Schrodinger}:
\begin{equation}\label{commutador.identity}
\Delta A \Delta B \geq \frac{1}{2}\left|\left\langle \left[\hat{A},\hat{B}\right]  \right\rangle\right|.
\end{equation}
Here, the statistical uncertainty $\Delta A $ is defined as follows:
\begin{equation}
\Delta A=\sqrt{\left(\Psi_{A}\cdot\Psi_{A}\right)},
\end{equation}
where $\Psi_{A}\equiv(\hat{A}-\left\langle A\right\rangle)\Psi$ for any normalized wave function $\Psi$ of the Hilbert space $\mathcal{H}$.

Classical statistical mechanics admits a similar operational interpretation of its uncertainty relations (\ref{canonical.uncertainty}). The basic idea is to admit the usual operational rule for the expectation values of physical observables:
\begin{equation}
\left\langle O \right\rangle=\int_{\mathcal{M}_{\theta}}\Phi(I|\theta)\hat{O}\Phi(I|\theta)dI.
\end{equation}
Of course, there is no needing to restrict the admissible operators $\hat{O}$ of thermodynamic observables to the class of \emph{hermitian operators} because of the probability amplitudes $\Phi(I|\theta)$ in classical statistical mechanics belong to a Hilbert space $\mathcal{H}_{R}$ of real functions. The only requirement is that the operators $\hat{O}$ of physical observables should be \emph{real operators}:
\begin{equation}
\hat{O}^{*}=\hat{O}.
\end{equation}
Considering the asymptotic probability amplitude (\ref{EP2}) associated with the correspondence principle of classical statistical mechanics, one can justify the introduction of the following operators:
\begin{equation}\label{op.class}
\hat{I}=I\mbox{ and }\hat{\mathbf{\eta}}=2k\frac{\partial}{\partial I}
\end{equation}
demanding the correspondence with thermodynamic variables in the limit $k\rightarrow 0$:
\begin{equation}
\hat{\mathbf{\eta}}\Phi(I|\theta)\sim\hat{\mathbf{\eta}}
e^{\frac{1}{2k}\mathcal{S}(I|\theta)}\rightarrow\mathbf{\eta}=\frac{\partial \mathcal{S}(I|\theta)}{\partial I}.
\end{equation}
The complementary character between the macroscopic observables $I$ and the restituting generalized forces $\mathbf{\eta}$ can be related to the fact that their respective operators $\hat{I}$ and $\hat{\mathbf{\eta}}$ do not commute:
\begin{equation}\label{commutador}
\left[\hat{I}^{j},\hat{\eta}_{i}\right]=-2k\delta^{j}_{i}.
\end{equation}
Of course, one cannot naively apply the Robertson-Schrodinger inequality (\ref{commutador.identity}) to this context because of the same one was derived in the framework of Hermitian operators only \cite{Robertson,Schrodinger}. For general operators $\hat{A}$ and $\hat{B}$, this inequality should be replaced by:
\begin{equation}\label{commutador.identity}
\Delta A \Delta B \geq \frac{1}{2}\left|\left\langle \hat{A}^{+}\hat{B}e^{i\phi}+\hat{B}^{+}\hat{A}e^{-i\phi}    \right\rangle\right|,
\end{equation}
where $\phi$ is an arbitrary phase. For the case of interest, the operators $\hat{I}^{i}$ are hermitian, $(\hat{I}^{i})^{+}=\hat{I}^{i}$, while the operators $\hat{\eta}_{i}$ are \emph{anti-hermitian}, $\hat{\eta}^{+}_{i}=-\hat{\eta}_{i}$. Considering that $\hat{A}$ is a hermitian operator, $\hat{A}^{+}=\hat{A}$, and $\hat{B}$ is \emph{anti-hermitian}, $\hat{B}^{+}=-\hat{A}$, inequality (\ref{commutador.identity}) is obtained when the phase parameter $\phi=0$. Thus, the inequality (\ref{commutador.identity}) can be also employed in the present situation, which leads to the uncertainty relations (\ref{canonical.uncertainty}) considering the commutation relation (\ref{commutador}).

\section{Final remarks}\label{Discussions}

The present discussion evidences that any physical theory with a statistical formulation should admit the existence of \emph{complementary quantities} as well as its interpretation in terms of \emph{non-commuting operators}. In particular, this claim has been shown for the case of classical statistical mechanics. Formally, quantum mechanics and classical statistical mechanics are two physical theories that share several analogies. In fact, foundations of classical statistical mechanics can be rephrased to introduce counterpart versions of some of orthodox principles of quantum mechanics.

As first glance, the underlying analogy between these physical theories is still uncomplete. For example, classical statistical mechanics can be formulated in terms of a real probability amplitude $\Phi(I|\theta)$ and some relevant operators of this theory (\ref{op.class}) are, indeed, linear operators. However, these arguments are insufficient to support that the probability amplitude $\Phi(I|\theta)$ obeys a counterpart of the \textbf{superposition principle} of quantum mechanics:
\begin{equation}
\Psi(\mathbf{q},t)=\sum_{i}a_{i}\Psi^{(i)}(\mathbf{q},t).
\end{equation}
In principle, the licitness of a superposition principle in terms of the real probability amplitude $\Phi(I|\theta)$ could be justified taking into account that classical statistical mechanics is an \emph{emergent physical theory}, that is, a macroscopic theory that is derived staring from microscopic physical laws. Physically speaking, it is reasonable to presuppose a certain hierarchical correspondence between the wave function $\Psi(\mathbf{q},t)$ and the real probability amplitude $\Phi(I|\theta)$. Perhaps, the analogy between classical statistical mechanics and quantum mechanics could be the manifestation of a deeper physics. For example, one can speculate that quantum and classical statistical mechanics correspondence principles are particular cases of a \textbf{unified asymptotic theorem}: \emph{The wave function of a closed (presumably nonlinear) system with large quantum numbers and large number of degrees of freedom (namely, in the limits $\hbar\rightarrow0$ and $k\rightarrow0$) asymptotically behaves as follows:}
\begin{equation}
\Psi(\mathbf{q},t)\sim\exp\left[\mathcal{S}(\mathbf{q},t)/2k+iS(\mathbf{q},t)/\hbar\right],
\end{equation}
\emph{where $S(\mathbf{q},t)$ is the classical action, while $\mathcal{S}(\mathbf{q},t)$ a microscopic counterpart of statistical entropy.} In principle, this type of hypothesis can be justified analyzing the correspondence between quantum and classical chaos \cite{Gutzwiller}.

On the other hand, Einstein postulate (\ref{EP}) has been always regarded as an \emph{exact expression} in classical fluctuation theory. Its present reinterpretation as the correspondence principle between classical statistical mechanics and thermodynamics implies, by itself, the restricted applicability of this hypothesis. The underlying analogy with quantum mechanics strongly suggests the existence in classical statistical mechanics of a counterpart of Schr\"{o}dinger equation:
\begin{equation}
i\hbar\frac{\partial}{\partial t}\Psi(\mathbf{q},t)=\hat{H}\Psi(\mathbf{q},t)
\end{equation}
describing the evolution of a closed system towards the thermodynamic equilibrium. As expected, such a dynamical equation should provide the correct probability amplitude $\Phi$ for any physical situation, which should correspond to the \emph{hydrodynamic transport equations} in the thermodynamic limit $k\rightarrow 0$. A crucial starting point in this generalization is the analogy between classical mechanics and thermodynamics \cite{Omohundro}. Taking into account the counterpart character of the classical action $S(\mathbf{q},t)$ and the entropy $\mathcal{S}(I|\theta)$, it is natural to expect that the entropy should play a central role in some type of \emph{variational approach} of hydrodynamic transport theory \cite{Sieniutycz}. Recently, Rajeev has shown that state equations describing the thermodynamic behavior of substance can be incorporated into the mathematical apparatus of Hamilton-Jacobi theory \cite{Rajeev}.

\emph{Acknowledgement}: Velazquez thanks the financial support of CONICYT/Programa Bicentenario de Ciencia y Tecnolog\'{\i}a PSD
\textbf{65} (Chilean agency).

\end{document}